\documentclass[11pt,titlepage,final,reqno,nosumlimits]{amsart}
\usepackage{epsfig}
\usepackage{graphicx}
\usepackage[latin1]{inputenc}

\usepackage{epsfig}
\usepackage{amssymb}
\newcommand{\diff}{{\rm d}}
\newcommand{\MM}{{\sc mm}}
\newcommand{\BFKL}{{\sc bfkl}}
\newcommand{\ART}{{\sc art}}
\newcommand{\ARP}{{\sc arpack}}
\newcommand{\FFTW}{{\sc fftw}}
\newcommand{\half}{\tfrac12}

\newcommand{\h}{\mathfrak{h}}

\numberwithin{equation}{section}

\begin{document}
  \begin{titlepage}
    \begin{flushright}
      {\small\sf UPRF-2004-10}
    \end{flushright}
    \vskip 1.in
    \begin{center}
      \textbf{BFKL, MM, Alpert-Rokhlin's transform, FFTW, ARPACK \& all that.}\\[3.em]
      \textbf{\large E.\ Onofri\footnote{\it Dipartimento di Fisica,
          Universit\`a di Parma, and {\small\sf I.N.F.N.}, Gruppo Collegato di
          Parma, 43100 Parma, Italy, {\sf
            \mbox{{onofri@unipr.it }}} }}\\[2.em]
    \end{center}
    
  \bigskip
  \bigskip
  \begin{center}
    \textbf{Abstract}
  \end{center}
  {\small The evolution equation for $q\,\bar q$ production introduced by
    Marchesini and Mueller posed some intriguing mathematical puzzles, both
    numerical and analytic. I give a detailed account of the numerical approach
    which led eventually to the exact solution. While part of the work was in
    fact along a wrong track, it turns out that some of the techniques involved
    are interesting in their own and applicable to many other problems, {\it
      i.e.\/} to the numerical study of Ricci flows.}
\end{titlepage}

\section{ Introduction }

Marchesini and Mueller \cite{Marchesini:2003nh,marchesini04:_exact_solut_bfkl}
introduced the following equation for the evolution of $q\,\bar q$ evolution in
QCD
\begin{equation}
\frac{  \partial\,u(\tau,\xi)}{\partial \tau} =
\int_0^1\frac{u(\tau,\eta\,\xi)/\eta-u(\tau,\xi)}{1-\eta}\,\diff\eta
+\int_{\xi}^1\frac{u(\tau,\xi/\eta)-u(\tau,\xi)}{1-\eta}\,\diff\eta
\end{equation}
where the unknown function $u(\tau,\xi)$ must vanish at $\xi\!=\!0$ to ensure
the convergence of the integrals involved. If we put $u(\xi)=\xi\,\psi(\xi)$
(boundary conditions are then taken into account if $\psi$ is bounded or at
least does not grow too rapidly at $\xi\to 0$)
\begin{equation}
  \label{eq:good}
  \frac{  \partial\,\psi(\tau,\xi)}{\partial \tau} \equiv 
  (K\psi)(\tau,\xi) = 
\int_0^2\,\frac{\diff \eta}{|\xi-\eta|}\,\big(\psi(\tau,\eta) -
  \min(1,\xi/\eta)\,\psi(\tau,\xi)\big)
\end{equation}
I shall refer to $K$ as the \MM\ operator.  We can easily discretize $K$ on a
lattice $\xi_n=na,\, a=2/(N+1)$
\begin{equation}\label{eq:2}
  (K\psi)_i = \sum_{j\ne i}\,\frac{\psi_j-\min(1,i/j)\,\psi_i}{|i-j|}\,.
\end{equation}
  
Any trace of $a$ disappears from the discrete equation which is a sign of the
scale invariance of the original equation: under $\xi\to\lambda\xi$ only the
endpoint changes, hence the result is insensitive to its actual value.  The
spectrum can be estimated numerically. By Richardson extrapolation from
dimension $32, 64, \ldots, 1024$ one gets for the spectrum of $K$

\begin{verbatim}
  E =
   2.4990 
   1.7993
   0.9604
   0.2179
  -0.3663
    ......
\end{verbatim}

$K$ has a negative spectrum except for a few positive eigenvalues, the largest
one dominates the evolution (all others are damped away).

Notice that if (accidentally, by mistake!), you ignore the ``${\rm
  min}(1,\xi/\eta)$'' factor in Eq.\eqref{eq:good} the spectrum comes out very
simple and, surprisingly enough, {\sl independent from $N$\/} to all available
figures:
{\small
\begin{verbatim}
     N = 32        64        128       256       
      -0.0000   -0.0000   -0.0000   -0.0000  
       2.0000    2.0000    2.0000    2.0000  
       3.0000    3.0000    3.0000    3.0000  
       3.6667    3.6667    3.6667    3.6667  
       4.1667    4.1667    4.1667    4.1667  
       4.5667    4.5667    4.5667    4.5667  
       4.9000    4.9000    4.9000    4.9000  
       5.1857    5.1857    5.1857    5.1857  
       5.4357    5.4357    5.4357    5.4357  
\end{verbatim}
}
What is the secret behind these numbers? Taking the differences
we get
{\small
\begin{verbatim}
                     2.0000              
                       1.0000            
                         0.6667          
                           0.5000        
                             0.4000      
                               0.3333    
                                 0.2857  
                                   0.2500
\end{verbatim}
}
\noindent
an easily recognizable sequence. Indeed the eigenvalues are given precisely by
twice the ``harmonic numbers'' $\{\h_n=\sum_{j=1}^n j^{-1}\;|\, n>0\}$ and
$\h_0\equiv 0$. This fact is actually an exact property of the modified integral
equation, both in its discretized form and on the continuum (a result which goes
back to the sixties \cite{tuck64}, see Appendix), the eigenvectors being {\sl
  Tchebyshev discrete polynomials \cite{Bateman55, nikiforov}\/} which
  converge to Legendre polynomials in the limit $N\to\infty$.

\section {Perturbation theory}
Following the hint of the previous Section, let us represent $K$ as the sum of
two terms and treat the problem by perturbation theory.
\begin{equation}
(K\psi)(\xi) = (K_0 \psi)(\xi) - \log(\xi)\;\psi(\xi)
\end{equation}
where
\begin{equation}
  (K_0\psi)(\xi) = \int_0^1 \diff\eta\frac{\psi(\eta)-\psi(\xi)}{|\xi-\eta|}
\end{equation}
$K_0$ is exactly diagonalizable, with eigenfunctions the Legendre polynomials
$P_n(2\xi-1)$ and eigenvalues proportional to the {\sl harmonic numbers\/}
$h_n=\sum_{j=1}^n j^{-1}$ (see Appendix A).  Second order perturbation theory
gives for the ground state $E_0 \approx 1.44754$ hence convergence appears to be
rather slow.  The usual methods to get high order coefficients are not
applicable here, since the matrix $\langle P_n|\log(\xi)|P_m\rangle$ is {\sl
  full\/}.

One can do better with a purely numerical approach as we discuss in the next
section  (the coefficients, we shall see, decrease only as $1/n\log(n)$, which
would require high orders in p.t. to get a meaningful result).

\section{Evolution}
I recyvled an old program which was used to study the renormalization group
equation of the non--linear sigma model (the ``sausage''
\cite{fateev93,Belardinelli:1994dq,Belardinelli:1995gt}). The equation is now
rather popular in the mathematical literature as the \emph{Ricci flow}. The idea
is to split the evolution of
\begin{equation}
  \frac{\partial\psi(\Delta,\xi)}{\partial\Delta} = K\,\psi(\Delta,\xi)
\end{equation}
into two steps 
\begin{equation}
  \begin{split}
  \psi_{\rm tmp}(\xi) &= \psi(\Delta,\xi)+ \tau K_0\,\psi(\Delta,\xi)\\
  \psi(\Delta+\tau,\xi) &= \psi_{\rm tmp}(\xi) - \tau
  \log(\xi)\,\psi(\Delta,\xi))
  \end{split}
\end{equation}
The first step is accomplished by going to the representation in terms
of Legendre polynomials ($\psi=\sum\psi_n\,P_n(2\xi-1)$) where $K_0$ is
diagonal. Coming back to the $\xi$-representation one executes the
second step. The program is implemented in {\sf matlab}.

\begin{figure}[ht] 
  \begin{center}
    \mbox{\epsfig{file=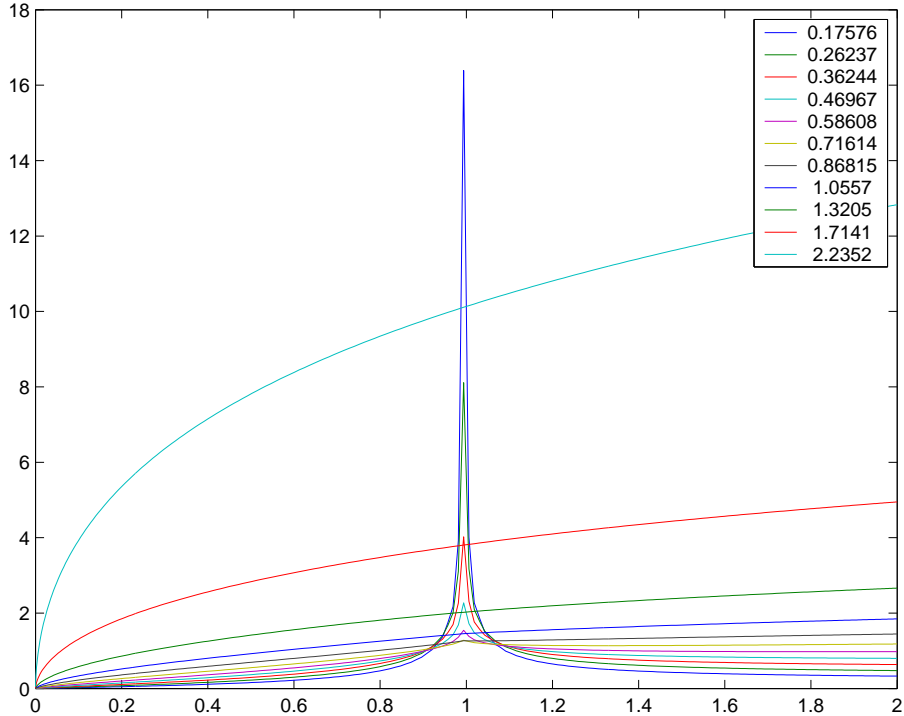,width=12.cm}}
    \caption{Evolution from $\xi_0=1$}
  \end{center}
\end{figure}

\begin{center}
  \begin{table}[ht]
    \begin{tabular}{|l|l|}\hline
      N & DLT  \\\hline
      32  & 2.030\\
      64  & 2.144\\
      128 & 2.235\\
      256 & 2.308\\
      512 & 2.368\\
      1024 & 2.417\\
      2048 & 2.457\\
      $\infty$ & 2.66 \\\hline
    \end{tabular}\vskip .1in
    \caption{$K$ ground state by the direct Legendre transform.}
  \end{table}\label{tab:discrete}
\end{center}

{\sl En passant} one can study the spectrum of $K$ within the same program. We
get the result of Tab.\ref{tab:discrete}, where the last line is obtained by
extrapolating in the variable $n^{-1/4}$, which appears at first sight as an
approximate scaling law (but see later on).  This value should be compared to
the approximate saddle point value $4\ln 2\approx 2.77$. The strong dependence
on the grid size is not surprising, since we have to deal with a singular scale
invariant integral operator.  The similar operator considered by Tuck
\cite{tuck02:_longit} is {\sl not} scale invariant and its cutoff dependence is
much flatter (i.e. better!).

\section{Alpert-Rokhlin's Fast Legendre Transform}
Due to the very slow convergence toward $\,N\!\!\to\!\infty\,$ it is desirable
to be able to calculate the spectrum with a high number of collocation points.
This is totally unfeasible with the direct method. The calculation with $n=4096$
required a work space of 1/2 GByte and going further was not possible on
available workstations. The way out is to apply some sparse matrix computational
tool which should be able to save memory and time. It was shown by Alpert and
Rokhlin \cite{alpert91} that it is possible to transform from a Legendre
expansion $\sum_{n\leqslant N} c_n\,P_n(x) $ to a Tchebyshev expansion
$\sum_{n\leqslant N} \tilde c_n\,T_n(x)$ in $O(N)$ time, even if the amount of
memory required may be rather large (at least $200\,N$ words). Since Tchebyshev
polynomials of the first kind are just trigonometric functions in disguise, the
Legendre transform is reduced to a combination of Alpert-Rokhlin's transform
(\ART) and cosine-Fourier-transform.  Using Alpert's implementation of \ART
\footnote{B.  Alpert very kindly provided us with his Fortran code.}  combined
with \FFTW\ in mode {\sf\small REDFT10/01} (see
Ref.\cite{fftw98})\footnote{http://www.fftw.org/\#documentation} we realized a
code essentially equivalent to the previous one but allowing for high
dimensional matrix representation of the operator. The ground state has been
computed for $N=2^k, k=6,7,\ldots,18$ giving the result of Tab.2. The difference
from the previous calculation is due to a different choice of discretization
grid (Gaussian integration points, {\sl i.e.} the roots of $P_N(x)$, in the
former case, Tchebyshev points, uniformly spaced in ${\rm acos}(x)$, in the
latter).  Notice that the results of the fast method anticipate those of the
direct method, that is the ``fast'' result at $N$ is close to the ``direct''
result at $2N$. In a sense the formal dimension of the real \FFTW\ ($2N$) is the
``true'' dimension.

For the technically--oriented reader we report the approximate timings of the
two algorithms in Appendix B.

The extrapolation at $N\!\!\to\!\infty\,$, assuming a power law scaling as
before, seems consistent, giving 2.6733 and 2.6692 (linear and quadratic fit
respectively) with the first method, 2.6661 and 2.6704 with the second. We would
conclude that the saddle point estimate $4\log(2)$ is correct within $4\%$.

\begin{center}
  \begin{table}[ht]\label{tab:fast}
    \begin{tabular}{|c|l|l|}\hline
      $\log_2N$ & DLT & \ART \\\hline
      5  & 2.0246& \hfil - \hfil \\
      6  & 2.1416& 2.2436\\
      7  & 2.2339& 2.3164\\
      8  & 2.3076& 2.3751\\
      9  & 2.3674& 2.4232\\
      10 & 2.4165& 2.4631\\
      11 & 2.4572& 2.4966\\
      12 & 2.4922& 2.5249\\
      13 & \hfil - \hfil& 2.5491\\
      14 & \hfil - \hfil& 2.5700\\
      15 & \hfil - \hfil& 2.5880\\
      16 & \hfil - \hfil& 2.6038\\               
      17 & \hfil - \hfil& 2.6176\\
      18 & \hfil - \hfil& 2.6298\\\hline 
    \end{tabular}  
\vskip .2in
    \caption{The ground state from Direct Legendre Transform (DLT) and
    from the fast algorithm (\ART+\FFTW)}
  \end{table}
\end{center}

\begin{figure}[ht] 
  \begin{center}\label{fig:conv}
    \mbox{\epsfig{file=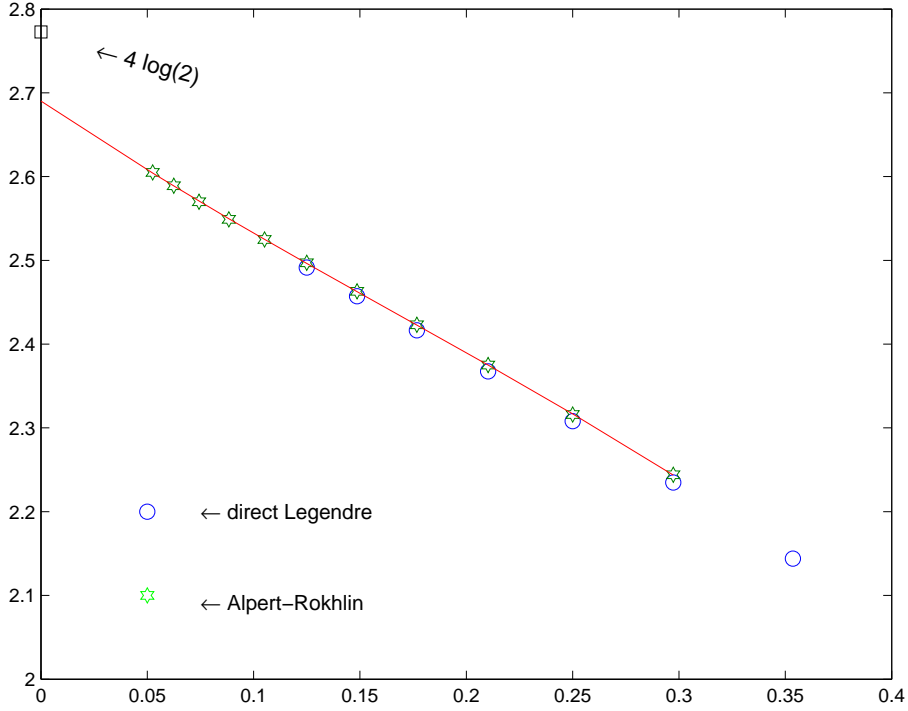,width=12.cm}}
    \caption{Continuum limit for the ground state with a power scaling law.}
  \end{center}
\end{figure}

\begin{figure}[ht] 
  \begin{center}\label{fig:log1}
    \mbox{\epsfig{file=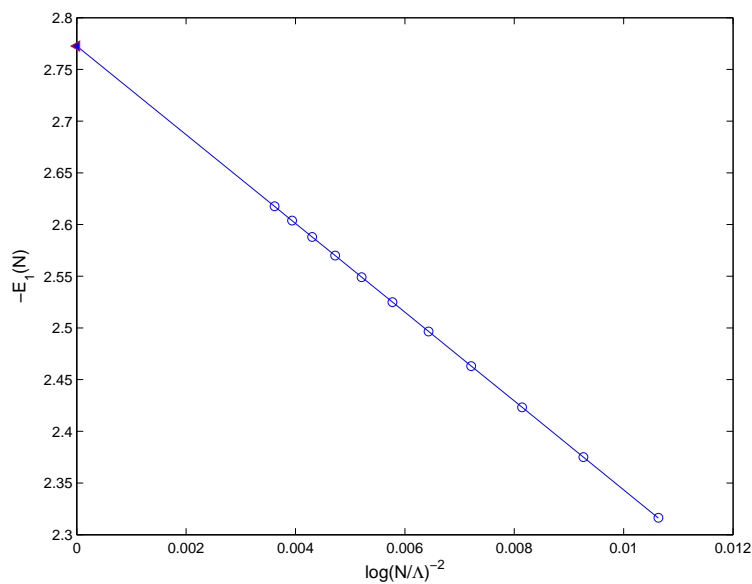,width=10.cm}}
    \caption{Continuum limit with the logarithmic  scaling law.}
  \end{center}
\end{figure}

\begin{figure}[ht] 
  \begin{center}\label{fig:logg}
    \mbox{\epsfig{file=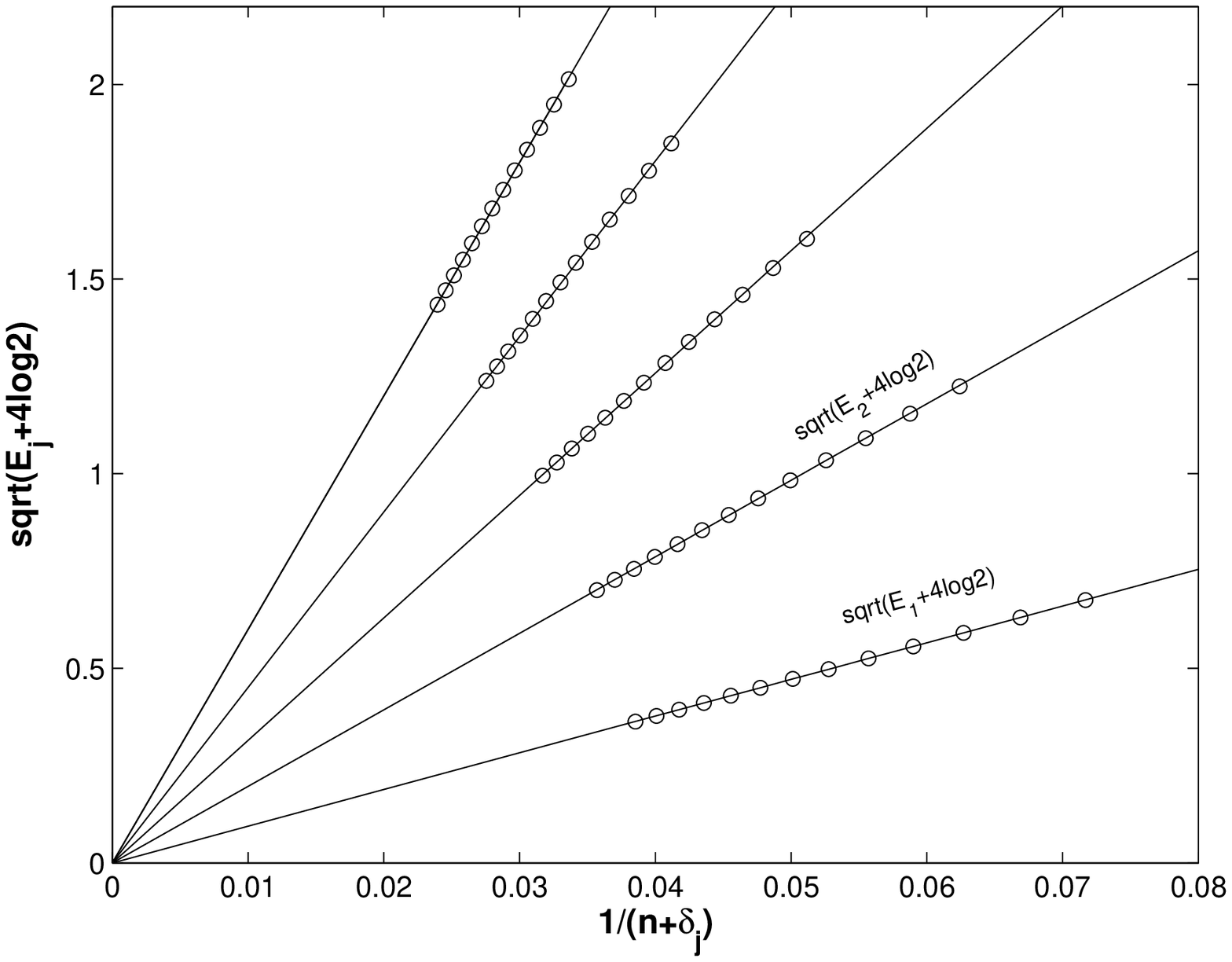,width=10.cm}}
    \caption{Extrapolation in log-scale shows the emergence of a continuous spectrum}
  \end{center}
\end{figure}

A totally different result is however hiding behind these figures. It must be
realized that the crucial point is to identify the correct $N$ dependence, since
this is going to make a big difference in the extrapolation at $N\to\infty$. A
careful analysis shows that a {\bf logarithmic scaling law} is much more
accurate than a power law. Looking for a fit of the kind $E(N) \approx E(\infty)
- C_1/\log(N/\Lambda) - C_2/\log(N/\Lambda)^2$ we get a very good interpolation
(the deviation is uniformly less than 1 part in $10^4$) and the value at
$N=\infty$ is compatible with $4\log(2)$ (within the same accuracy). According
to this idea we should conclude that, surprisingly enough, the saddle point
value is actually exact (see Figg.~3,4).  In the case studied by Tuck we find a
much steeper scaling law of the kind
$$
E(N) = E_1(\infty) + C N^{-2}\log(N)^{-1}
$$
as shown in Fig.~5.

\begin{figure}[ht] 
  \begin{center}
    \mbox{\epsfig{file=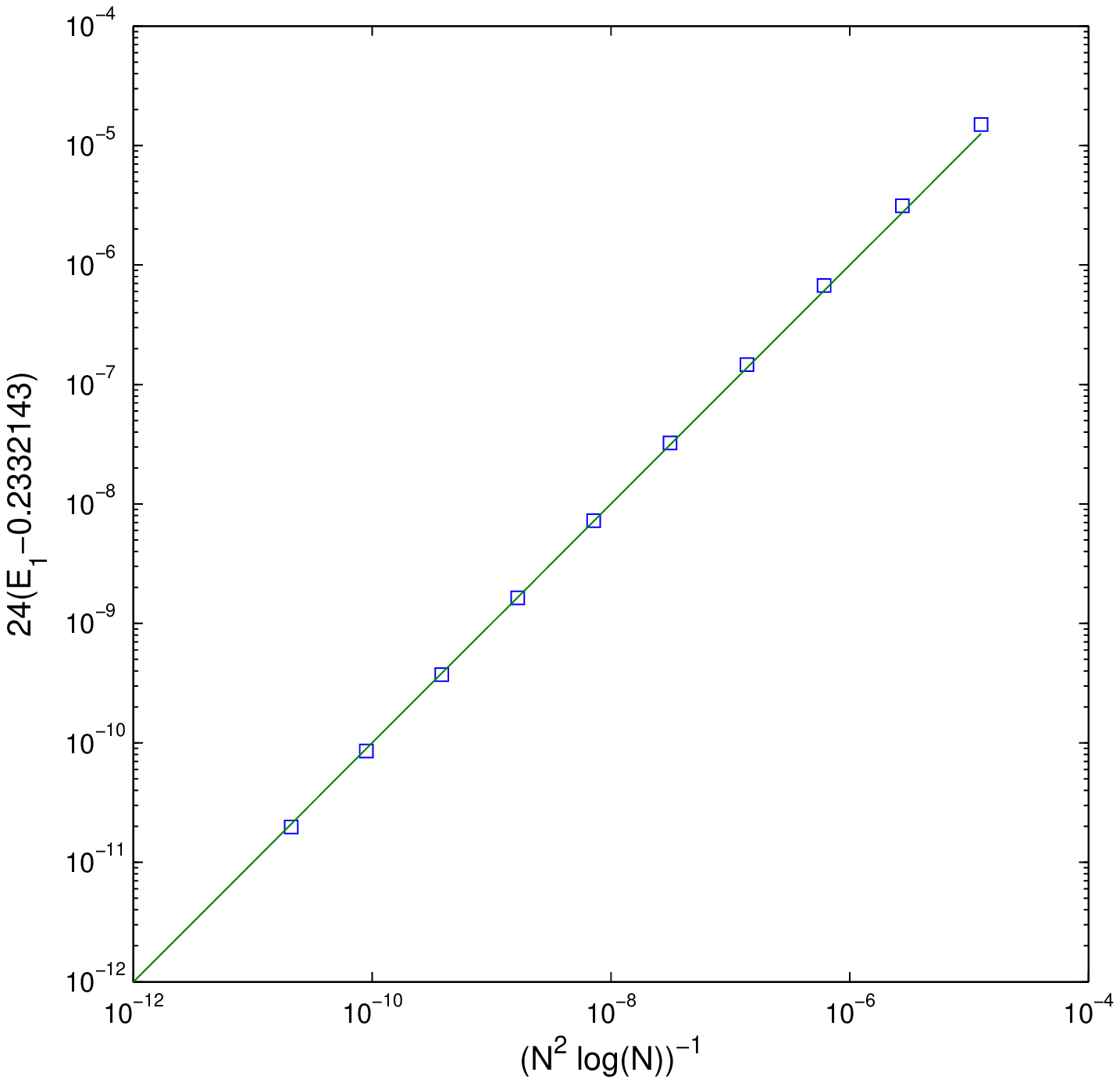,width=10.cm}}
    \caption{Continuum limit with the modified  scaling law, Tuck's case.}
  \end{center}
\end{figure}

It has been realized \cite{marchesini04:_exact_solut_bfkl} that the picture is
simply due to the different character of the spectrum: continuous for \MM\ and
discrete for Tuck's operators.  This fact is made absolutely transparent by
adopting a different representation of the operator $K$:
\begin{equation}
  \begin{split}
    (K\,\phi)(x) &=(K_0\,\phi)(x)+ 2\log(1+e^{-x/2})\,\phi(x) -4\log2\,\phi(x)\\
    (K_0\,\phi)(x) &\equiv
    \int_0^\infty\frac{\phi(x)-\phi(y)}{2\sinh|\half(x-y)|}\,\diff y\,.
      \end{split}
\end{equation}
The integral operator $K_0$ is almost local and it is not very different from a
kinetic term. If we consider a wave-function with support in a region far from
the origin, the operator reduces to
\begin{equation}
  K_0\phi(x) \approx
\int_{-\infty}^\infty\frac{\phi(x)-\phi(y)}{2\sinh|\half(x-y)|}\,\diff y 
\end{equation}
which is diagonal in Fourier space with eigenvalue $-\chi(p)$, the subtracted
Lipatov function, explicitly given by
$\chi(p)=2(\psi(1)-\Re(\psi(\half+ip)))-4\log2$, $\psi$ being the logarithmic
derivative of the $\Gamma$ function.  $K_0$ is well-known, not necessarily in
this form, as the \BFKL\ operator \cite{r.k.ellis98:_qcd}.

It has been realized that the representation introduced here is also more
convenient to allow a numerical study of the evolution in the case of \MM, while
this is not the case for Tuck's equation. Essentially the dominant eigenvalue
$4\log2$ is already built--in, while in the representation of
Sec.~\ref{tab:discrete} this value can only be obtained by extrapolating at very
large matrix dimensions (see Tab.~\ref{tab:fast}).

To make the difference between \MM\ and \BFKL\  operators more explicit, it would
be desirable to be able to apply the method of images (which is usually employed
with local differential operators) to get rid of the boundary. However no simple
boundary condition seems to be appropriate. Actually by solving the eigenvalue
equation by standard linear algorithms ({\sf\small matlab}'s {\sl eig} routine)
one finds that the eigenvectors are essentially shifted trigonometric functions,
{\it i.e.\/}, far from the boundary, 
$
  \phi_k(x) \approx \sin(k x + \delta(k))
$
. By switching $V(x)$ on and off, we can easily check that
the behavior at $x=0$ is strongly influenced by $V(x)$. 

The {\sl phase shift} $\delta(k)$ is particularly interesting. For example, the
asymptotic behaviour of $\phi(\Delta,x)$ at large $\Delta$ is strongly
influenced by it. This fact is well-known in the theory of potential scattering
in quantum mechanics. While the general setup here is quite different,
nonetheless there are remarkable analogies which give useful guidelines. For
example the vanishing of $\delta$ at $k=0$ is a signal of the absence of bound
states ({\sl Levinson's theorem\/}), were we able to extend the theorem to this
context. Details can be found in \cite{marchesini04:_exact_solut_bfkl}.

The ``unbounded'' representation helps in understanding what really goes wrong
with the initial approach based on the Legendre basis.  Introducing a finite box
of side $L$ ($0\leqslant x \leqslant L$) the energy spectrum is discretized and
at low energy it is given by $\propto(n\pi/L)^2$. In the Legendre expansion of
the previous section all Gaussian points are confined to $x \lesssim L =
2\,\log(N)$.  This fact explains the logarithmic scaling law depicted in Fig.~3.
Also, since a good description of the evolution at large $\tau$ requires $L\gg
100$, this cannot be explored through the Legendre expansion.

\section{Further developments}
Recent developments pushed our understanding of the problem to a higher level.
A precise characterization of the time--dependent solution of \MM\ equation was
developed by a perturbative technique which can be pushed to all orders and
allows for a full resummation \cite{marchesini04:_exact_solut_bfkl}. An exact
form for the phase shift and the continuum eigenfunctions has been derived. A
rigorous proof on purely algebraic grounds, thus avoiding a delicate problem of
resummation, has been later found, thanks to an idea of V.\ A.\ Fateev
\cite{fateev04:_exact}.

There exists another representation of the integral operator which avoids the
presence of a boundary. Thanks to the intrinsic scale invariance, the equation
can be remapped on the whole of $\mathbb R$ by setting $\exp(x)=\exp(x')+1$,
which leaves the kernel invariant and only modifies the potential. In this
representation we may apply a spectral algorithm to the evolution equation
simply based on Fourier transform, more economic than the combined \ART+\FFTW.
This will be left as a homework.

\section{Conclusions and outlook}
The integral equation introduced by Marchesini and Mueller is deeply related to
another problem in mathematical physics studied by E. Tuck fourty years ago.
The connection to Tuck's equation was used to analyze \MM\ operator's spectral
properties by an efficient (sparse) matrix computation, based on Alpert-Rokhlin
transform, \FFTW\ and \ARP.  This analysis suggests that the spectrum is
continuous with endpoint $4\log2$. A second representation of the integral
operator makes the spectral properties more transparent and lends itself to an
easier algorithmic implementation which allows to evaluate the evolution at
large $\Delta$.

The application of \ART\ to the renormalization group equation for $O(3)$
$\sigma$-model may be useful to achieve greater accuracy than allowed from the
direct transform \cite{Belardinelli:1995gt}. More generally, the application of
a full group theoretical transform without axial symmetry, will make it possible
to explore the $O(3)$ Ricci flow in full detail. 

\section*{Acknowledgments}
I would like to thank warmly Prof. {\bf G. Altarelli}, chairman, and the staff
of the Theory Division of CERN, for the kind hospitality he enjoyed while most
of this work was done. I warmly thank {\bf B.\ K.\ Alpert} for making his
Fortran implementation of \ART\ available, {\bf G.\ E.\ Andrews} and {\bf R.\ 
  A.\ Askey} for pointing out to us Tuck's relevant paper and {\bf E.\ O.\ Tuck}
for very valuable correspondence.  {\bf S.\ Shaw}'s header files, available on
the WEB, proved to be very helpful for an \ARP\ beginner.  This work would have
been impossible without the searching capabilities of {\sf Google}.  But, above
all, I'm indebted to my dear friends {\bf G.\ Marchesini} and {\bf V.\ A.\ 
  Fateev}, for constantly providing new ideas, suggestions and insight.

\section*{Appendix A}
\noindent

Here is a proof for the discrete form of Tuck's operator.  For the original
problem see \cite{tuck64,fateev04:_exact}.

Let 
\begin{equation}
  (K_0^{(N)}\,v)_i = \sideset{}{'}\sum_{0\leqslant j\leqslant N}\, \frac{v_j - v_i}{|i-j|}\,.
\end{equation}
Let's apply the matrix $K_0^{(N)}$ to the vector $v_i^{(\ell)}=i^\ell$. We have
  \begin{equation}
    \begin{split}
      (K_0^{(N)}\,v^{(\ell)})_i &= \sideset{}{'}\sum_{0\leqslant j\leqslant N}\, \frac{j^\ell - i^\ell}{|i-j|}\\
      &=\left(-\sum_{j=0}^{i-1}+\sum_{j=i+1}^N\right) \sum_{k=0}^{\ell-1}\,i^{\ell-k-1}\,j^k\\
      &=
      \sum_{k=0}^{\ell-1}\,i^{\ell-k-1} \left(-\frac {2\, i^{k+1}}{k+1}+O(i^k)\right)\\
      &= -2\,\h_\ell\,v_i^{(\ell)} + \sum_{n<\ell}\,c_n\,v_i^{(n)}
    \end{split}
  \end{equation}
  where we used the Euler-MacLaurin summation formula for $\sum_j\,j^k$, and the
  constants $\{c_n\}$ are calculable but unnecessary.  This proves that $K^{(N)}
  \, v^{(\ell)}$ is contained in the linear span of $[v^{(0)}, v^{(1)}, ...,
  v^{(\ell)}]$. Since $K_0(N)$ is symmetric, it is diagonable, its eigenvectors
  are orthogonal, hence they are given by the orthogonal discrete polynomials
  with respect to the uniform weight on the set $[0, 1, 2, \ldots, N]$. The
  eigenvalues can be read off the coefficient of $v^{(\ell)}$ in the expansion
  of $K_0^{(N)}\,v^{(\ell)}$. The explicit form of the eigenvectors is given by
  Tchebyshev polynomials of a discrete variable
  \cite{abramowitz65:_handb_mathem_funct}.

\section*{Appendix B}
We give here some technical details about the algorithms which we have applied
in the paper.  The {\sl direct method\/} consists in building the table of
Legendre polynomials $\{P_n(x_j^{(N)}),\, n=0,1,\ldots,\mbox{N-1}\}$ at the
Gauss points, i.e. at the roots of $P_N(x)$. The technique, exploiting the
recurrence relation of orthogonal polynomials, is due to Golub and Welsch (see
Ref.\cite{golub69}).  To find the spectrum of $K$ we simply define $K_0$ to be
diagonal in the basis $\{P_n(x)\}$ with eigenvalues $-2 \h_n$.  The matlab
routine ``eig'' is then invoked. The singularity of the logarithm at the
boundary is avoided because the zeros of the polynomials are all internal at the
interval $[-1,1]$. The matrix representation of the free part $K_0$ is exact,
since the Gauss quadrature formula is exact on polynomials of low degree.  In
finite precision arithmetic $K_0$ is affected by the accumulation of truncation
errors, yielding an error of order $10^{-13}$ on its spectrum, which is rather
irrelevant. The method is presently feasible for dimension less than 4000 and it
has the advantage that $N$ can be any integer, not necessarily a power of 2.

The method based on \ART\ makes use of the expansion on Tchebyshev's polynomials
of the first kind $T_n(x)=\cos(n\, {\rm arccos}(x))$.  Again the Gauss points are
interior at the interval and the singularity is avoided. The real DFT of kind
$REDFT10$ precisely makes use of this grid of points.  Even if the
Gauss-Tchebyshev integration is exact for polynomials of low degree, still a
problem arises, namely that $K_0$ is a symmetric operator with respect to the
Lebesgue measure whereas Tchebyshev's polynomials are orthogonal with respect to
a different measure. It turns out that to restore symmetry we have to deal with
$\tilde K_0 = (1-x)^{1/4}\,K_0(1-x)^{-1/4}$, hence we are outside any polynomial
subspace and this introduces a systematic error making $\tilde K_0$ only
approximately symmetric. This has been checked after realizing that the spectrum
of $K_0$ considered as a symmetric operator contains substantial error, up to
10\%.  The strategy we adopt is therefore to relax the symmetry condition on
$\tilde K_0$ and compute the spectrum with a version of the Arnoldi algorithm
which allows to deal with non-symmetric operators provided by \ARP
\cite{arpack}. The combination \ART+\FFTW+\ARP (this latter offers the routine
{\sf znaupd} which applies to general non--symmetric complex matrices) turns out
to be again in the game with an accuracy on the spectrum of $K_0$ comparable if
not superior to the direct method. For example at $N=64$ we find

\begin{verbatim}
E=[
        1.00047386511033e-18
        2.00000000000000e+00
        2.99999999999999e+00
        3.66666666666666e+00
        4.16666666666666e+00
        4.56666666666665e+00
        4.90000000000001e+00
        5.18571428571429e+00
];
\end{verbatim}
and the reader can verify by herself that the error is only at the last decimal
place. A further (marginal) improvement will be achieved using the
real--non-symmetric routine {\sf dnaupd}; we use the complex version because it
was already implemented as a {\sf C++} header by Scott Shaw.

Let us now comment upon performance.  The first methods grows in time and memory
rather quickly ($O(N^2)$ in memory and $O(N^3)$ in execution time. The fast one
is much less memory greedy.  Notice that in the largest case examined,
$N=2^{18}$, the program requires slightly less than half a Gigabyte of memory,
half of which simply to allocate \ART's working array ($200N$ words). By contrast
the direct method requires $\approx 450$ MBy at $N=4096$ and it would grow to
the order of 6 TBytes at $N=2^{18}$ while the execution would require 50 years
at the present cpu speed.

Tables 3,4 report the execution times on a pentium III with
clock at 1.13 GHz and on a Xeon with clock at 2.8 GHz, respectively,
using Matlab v.6.5.  Execution times for the ``fast'' algorithm
are inclusive of the preconditioning (we select the initial
vector by executing a number of Trotter steps). As it is rather
clear from the table, the execution time grows as expected as
$O(N\log(N)$, with some expected deviations when the system
switches to virtual memory.

\begin{figure}[ht] 
  \begin{center}\label{fig:bench}
    \mbox{\epsfig{file=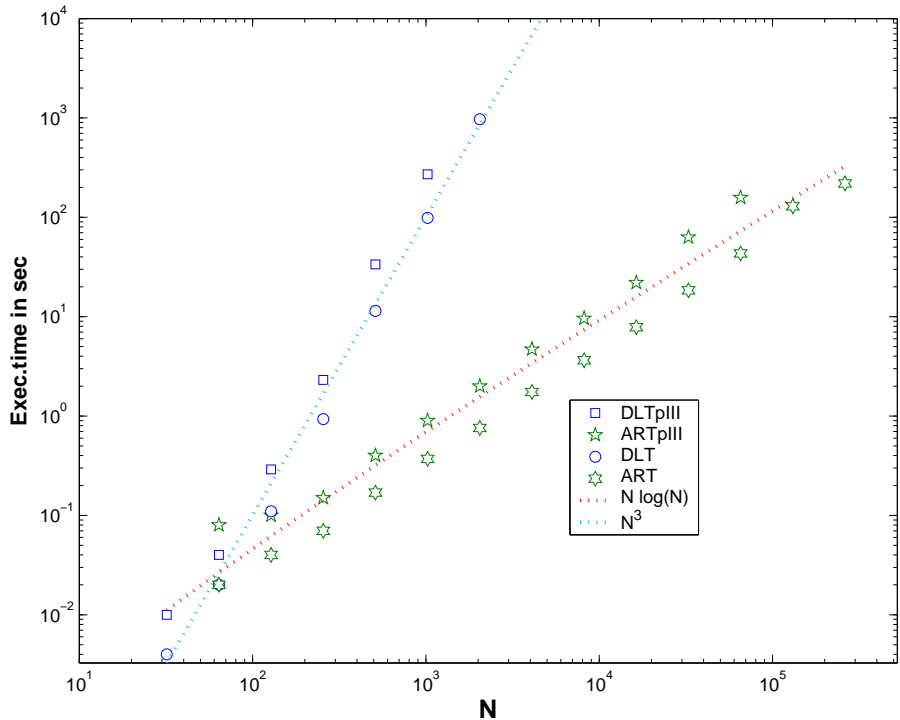,width=12.cm}}
    \caption{Execution times for the Direct Legendre transform and Alpert-Rokhlin transform.}
  \end{center}
\end{figure}

\begin{center}
  \begin{table}[ht]
    \begin{tabular}{|c|l|l|}\hline
      $\log_2N$ & DLT & \ART \\\hline
      5 &0.01  & \hfil - \hfil \\
      6 &0.04  & 0.08 \\
      7 &0.29  & 0.10 \\
      8 &2.31  & 0.15 \\
      9 &33.6  & 0.4 \\
      10&271.  & 0.9 \\
      11 &  \hfil - \hfil &2.0 \\
      12 &  \hfil - \hfil &4.7 \\
      13 &  \hfil - \hfil &9.6 \\
      14 &  \hfil - \hfil &22 \\
      15 &  \hfil - \hfil &63 \\
      16 &  \hfil - \hfil &158 \\\hline
    \end{tabular}  
\vskip .2in
    \caption{Timings (sec) (DLT) direct method  and
    for the fast algorithm (\ART+\FFTW); Pentium-III @ 1.13 GHz}
  \end{table}\label{tab:timP}
\end{center}

\begin{center}
  \begin{table}[ht]
    \begin{tabular}{|c|l|l|}\hline
      $\log_2N$ & DLT & \ART \\\hline
      5 &0.004  & \hfil - \hfil \\
      6 &0.02  & 0.02 \\
      7 &0.11  & 0.04 \\
      8 &0.93  & 0.07 \\
      9 &11.44  & 0.17 \\
      10& 98.6  & 0.37 \\
      11 & 970. & 0.76 \\
      12 &  \hfil - \hfil &1.75 \\
      13 &  \hfil - \hfil &3.64 \\
      14 &  \hfil - \hfil &7.84  \\
      15 &  \hfil - \hfil &18.4 \\
      16 &  \hfil - \hfil &43.4 \\
      17 &  \hfil - \hfil &129.\\
      18 &  \hfil - \hfil &219. \\\hline
    \end{tabular}  
\vskip .2in
    \caption{Timings (sec) (DLT) direct method  and
    for the fast algorithm (\ART+\FFTW); Xeon @ 2.8 GHz}
  \end{table}\label{tab:timX}
\end{center}

\bibliographystyle{unsrt}
\bibliography{uprf0410}

\end{document}